\documentclass{PoS} 
\pdfoutput=1

\usepackage{graphicx}
\usepackage{amssymb}
\usepackage{amsmath}
\usepackage{xcolor}
\usepackage{listings}
\usepackage{booktabs}
\usepackage{siunitx}
\usepackage{subcaption}
\usepackage{xspace}

\newcommand\gridarray{Gridarray\xspace}
    
\title{Lattice QCD on a novel vector architecture}

\ShortTitle{Lattice QCD on a novel vector architecture}

\author{Benjamin Huth, Nils Meyer, \speaker{Tilo Wettig}\\
    Department of Physics, University of Regensburg, 93040 Regensburg, Germany\\
    E-mail: \email{benjamin.huth@ur.de}, \email{nils.meyer@ur.de}, \email{tilo.wettig@ur.de}}

\abstract{
The SX-Aurora TSUBASA PCIe accelerator card is the newest model of NEC's SX architecture
family.  Its multi-core vector processor features a vector length of 16 kbits and
interfaces with up to 48~GB of HBM2 memory in the current models, available since 2018.
The compute performance is up to 2.45~\mbox{TFlop/s} peak in double precision, and the
memory throughput is up to 1.2~TB/s peak.
New models with improved performance characteristics are announced for the near future.
In this contribution we discuss key aspects of the SX-Aurora and describe how we
enabled the architecture in the Grid Lattice QCD framework.
}

\FullConference{The 37th Annual International Symposium on Lattice Field Theory - Lattice2019\\
		16-22 June 2019\\
		Wuhan, China}

\begin{document}

\section{Introduction}

Grid \cite{Boyle:2015tjk} is a modern Lattice QCD framework targeting parallel architectures.
Architecture-specific code is
confined to a few header files.  The CPU implementations use compiler built-in functions
(a.k.a.\ intrinsics) and assembly.  There is also a generic, architecture-independent
implementation based on C/C++ that relies on auto-vectorization.

Mainline Grid is limited to a vector register size of at most 512 bits. Here, we consider a new architecture with 16-kbit vector registers. We describe how we modified Grid to enable larger vector lengths and present initial performance benchmarks.

\section{NEC SX-Aurora TSUBASA}

\subsection{Overview}

\begin{figure}\vspace*{-7mm}
\begin{minipage}[t]{0.48\linewidth}
        \begin{center}
            \includegraphics[width=0.95\linewidth]{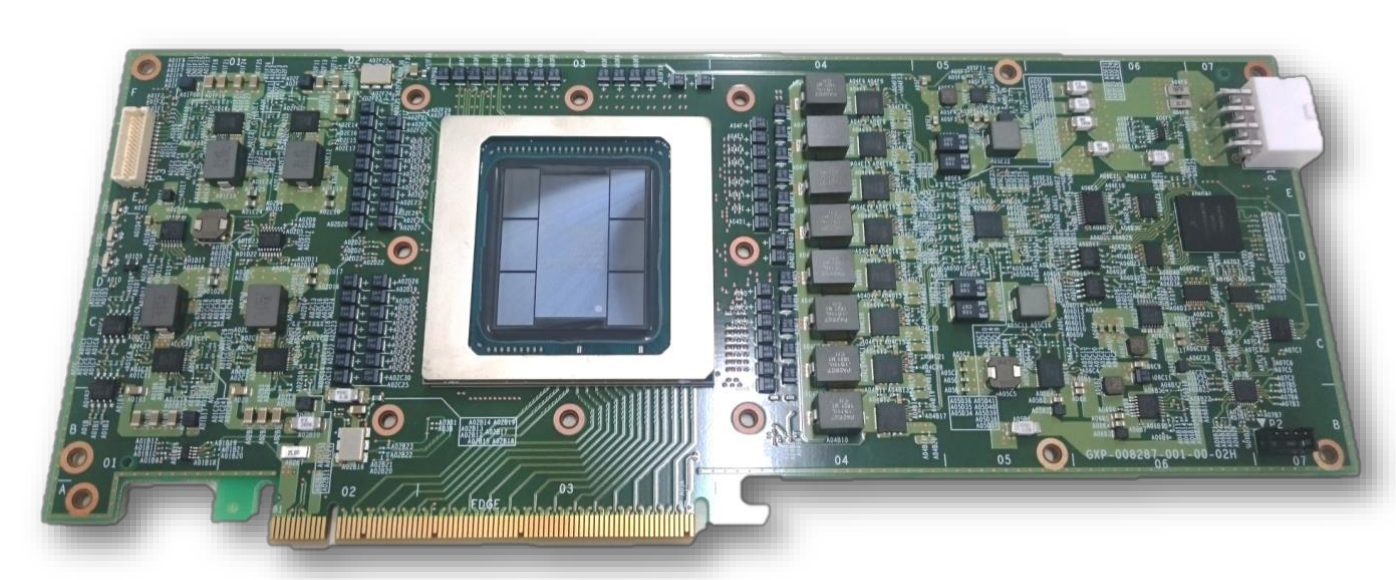}\\
            \caption{\label{aurora:card}NEC SX-Aurora TSUBASA 
            PCIe accelerator card (type 10).  Picture published with permission
            from NEC, \copyright\;by NEC.}
        \end{center}
\end{minipage}
\hfill
\begin{minipage}[t]{0.48\linewidth}
        \begin{center}
            \includegraphics[width=0.95\linewidth]{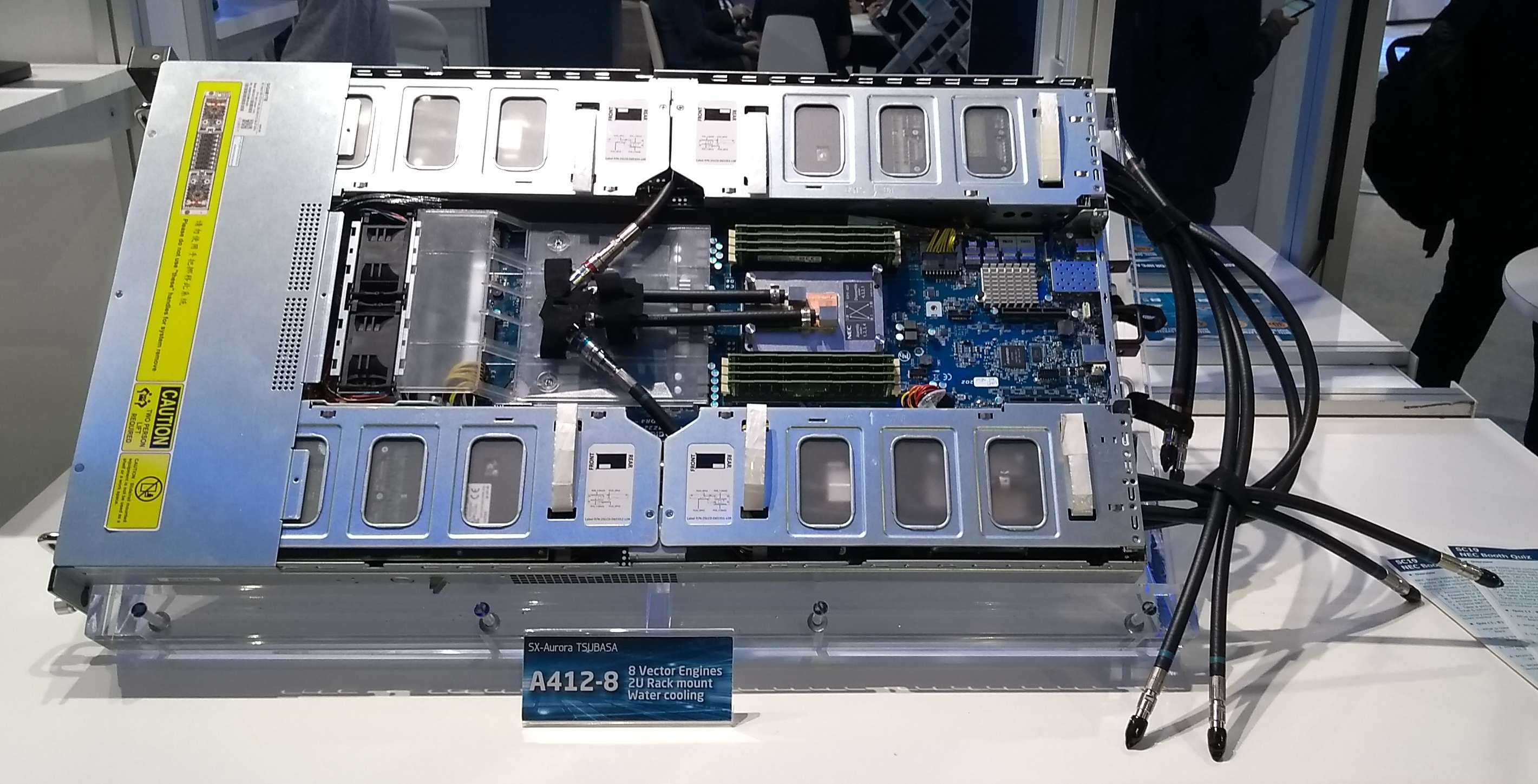}\\
            \caption{\label{aurora:a412}Liquid-cooled NEC A412-8 server
            presented at SC 19,  featuring 8 SX-Aurora cards of
            novel type 10E attached to a single-socket AMD Rome host
            CPU and fitting in 2U.}
        \end{center}
\end{minipage}
\end{figure}

\begin{table}[b]
    \begin{center}\small
    \begin{tabular}{l|rrr}
        Vector engine model\phantom{+++++++++++} & \quad Type 10A & \quad Type 10B  & \quad Type 10C \\\hline\hline
        Clock frequency [GHz]   & 1.6       & 1.4       & 1.4 \\
        SP/DP peak performance [TFlop/s] & 4.91/2.45 & 4.30/2.15    & 4.30/2.15 \\
        HBM2 capacity [GB]      & 48        & 48        & 24 \\
        Memory throughput [TB/s]& 1.20      & 1.20      & 0.75
    \end{tabular}
    \caption{\label{aurora:models}NEC SX-Aurora TSUBASA type 10 models.}
    \end{center}
\end{table}

The SX-Aurora TSUBASA, also called vector engine (VE), is the newest member of
NEC's SX series \cite{nec}.  In contrast to former vector supercomputer
architectures, the SX-Aurora is designed as an accelerator card, see Fig.~\ref{aurora:card}.
At present it is available with PCIe Gen3 x16 interconnect (VE type 10).  The
accelerator hosts a vector processor with 8 cores.  The card ships in 3 models,
which we list in Table~\ref{aurora:models}.  For instance, the type 10A flagship
model clocks at 1.6~GHz and delivers 2.45~TFlop/s DP peak.
The High Bandwidth Memory (HBM2) capacity is 48~GB with a throughput of 1.2~TB/s peak.
Improved accelerator models with higher main memory throughput
(type 10E, type 20) and 10 cores (type 20) have been announced
\cite{momose:rwth, wikichipfuse:roadmap}.

Multiple SX-Aurora platforms are available, including workstation, rack-mounted
server and supercomputer \cite{nec}.  Up to 64 vector engines interconnected by
InfiniBand fit into one A500 rack, delivering 157~TFlop/s DP peak.  In
Fig.~\ref{aurora:a412} we show the novel A412-8 server presented at SC 19.

\subsection{Vector engine architecture}

\begin{figure}
    \begin{center}
        \includegraphics[width=1.0\linewidth]{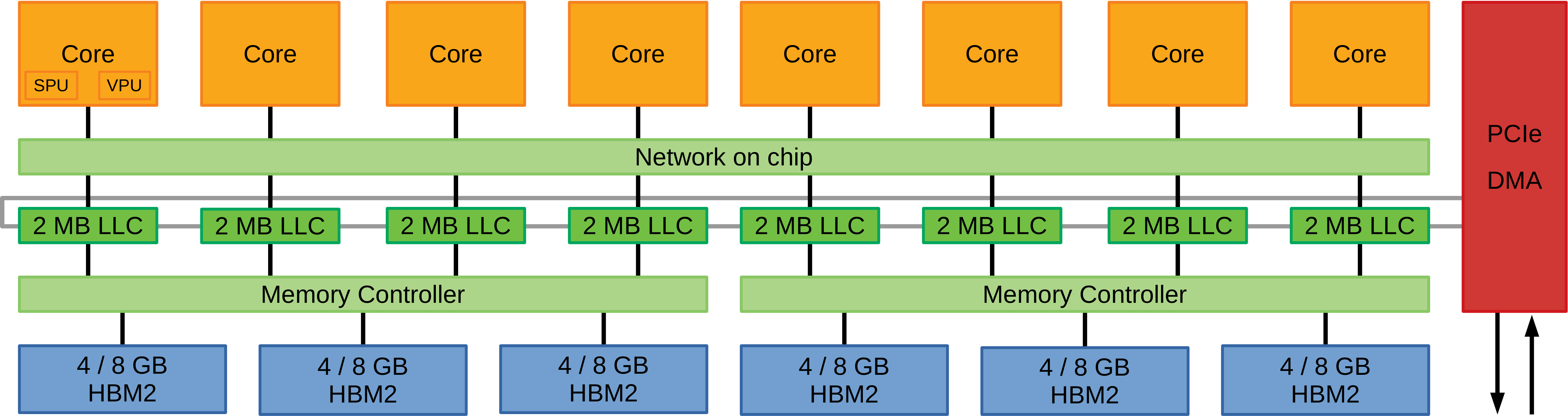}
        \caption{\label{aurora:arch}High-level architecture of the SX-Aurora
        type 10.}
    \end{center}
\end{figure}

The high-level architecture of the SX-Aurora type 10 is shown in
Fig.~\ref{aurora:arch} \cite{Yamada:hc18, wikichip}.
The chip contains 8 identical single-thread
out-of-order cores.  Each core comprises a
scalar processing unit (SPU) with 32 kB L1 cache and 256 kB L2 cache
as well as a vector processing unit (VPU).
The VPU processes
(optionally masked) 16-kbit vector registers (corresponding to 256 real DP numbers) in 8 chunks of 2 kbits each.

There are 8 blocks of 2 MB last-level cache (LLC) connected by a 2d network on chip.
The VPUs directly access this (coherent) LLC.
Two groups of 4 LLCs are connected to one memory controller each.
Every controller addresses 3 stacks of HBM2.
A ring bus interconnects the LLCs and allows for direct memory
access (DMA) and PCIe traffic.

\subsection{Programming models and software stack}

The SX-Aurora does not run an operating system.  It supports multiple modes
of operation described in \cite{Komatsu:sc18}.  Most attractive to us is the
VE execution mode: all application code is run on the accelerator, while system calls
are directed to and executed on the host CPU.

Two C/C++ VE compilers by NEC are available: ncc (closed source) and clang/LLVM VE (open source).
The latter is still in an early development stage.
Both compilers support OpenMP for thread-level parallelization but differ in how they vectorize.
ncc exclusively relies on auto-vectorization and does not support intrinsics, while clang/LLVM VE currently does not support auto-vectorization and relies on intrinsics instead.

The NEC software stack also includes support for MPI, debugging, profiling (ncc only) and optimized math libraries, e.g., BLAS and LAPACK.

\section{Grid on the NEC SX-Aurora}

\subsection{Enabling larger vector lengths in Grid}
Grid decomposes the lattice into one-dimensional arrays
(which we shall call \gridarray{}s in the following) that usually have the same size as the architecture's vector registers.
Operations on \gridarray{}s can achieve 100\% SIMD efficiency.

\begin{figure}[t]
    \begin{minipage}[c]{0.22\linewidth}
        \centering
        \includegraphics[width=\linewidth]{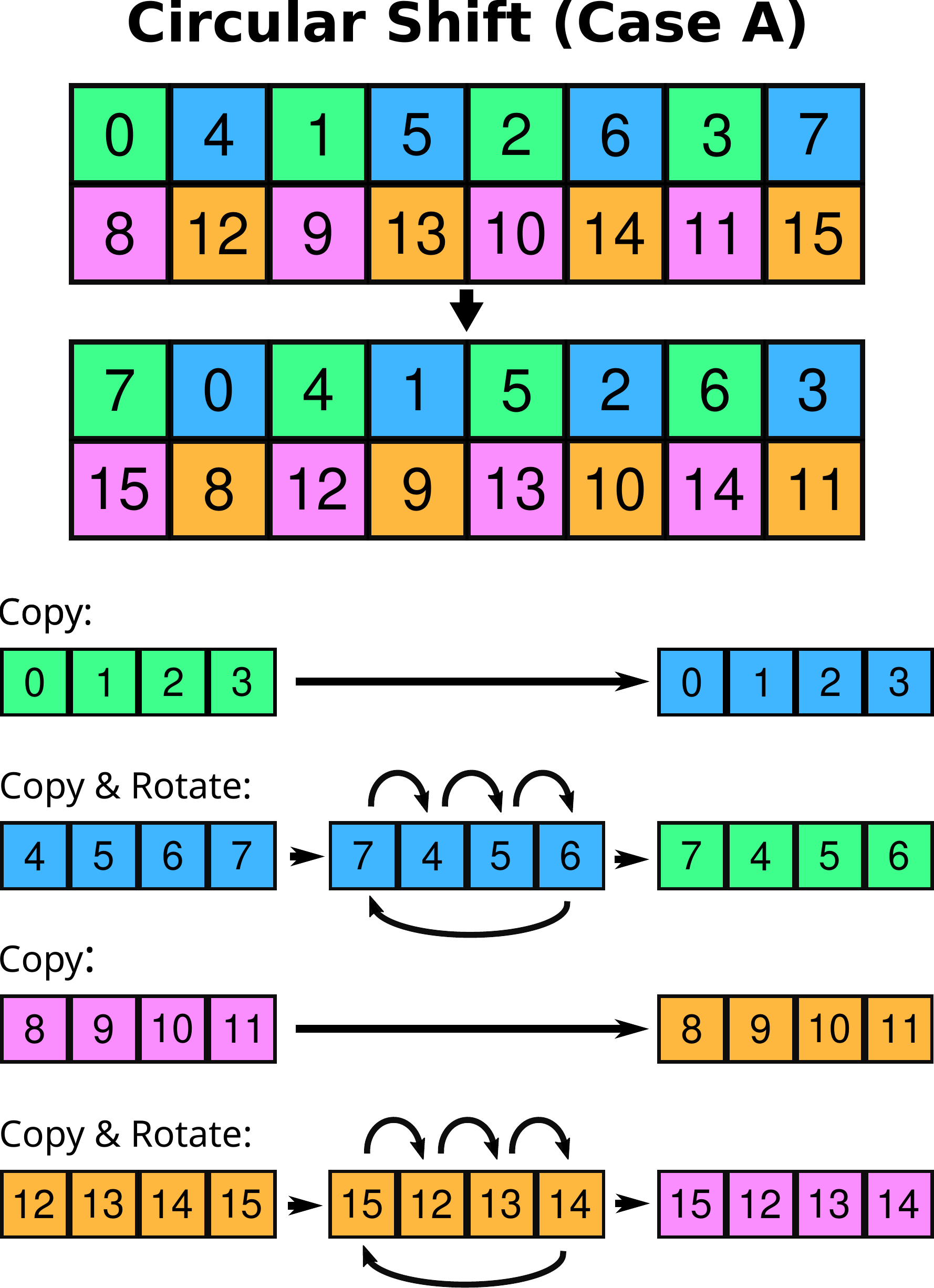}
    \end{minipage}
    \hfill
    \begin{minipage}[c]{0.22\linewidth}
        \centering
        \includegraphics[width=\linewidth]{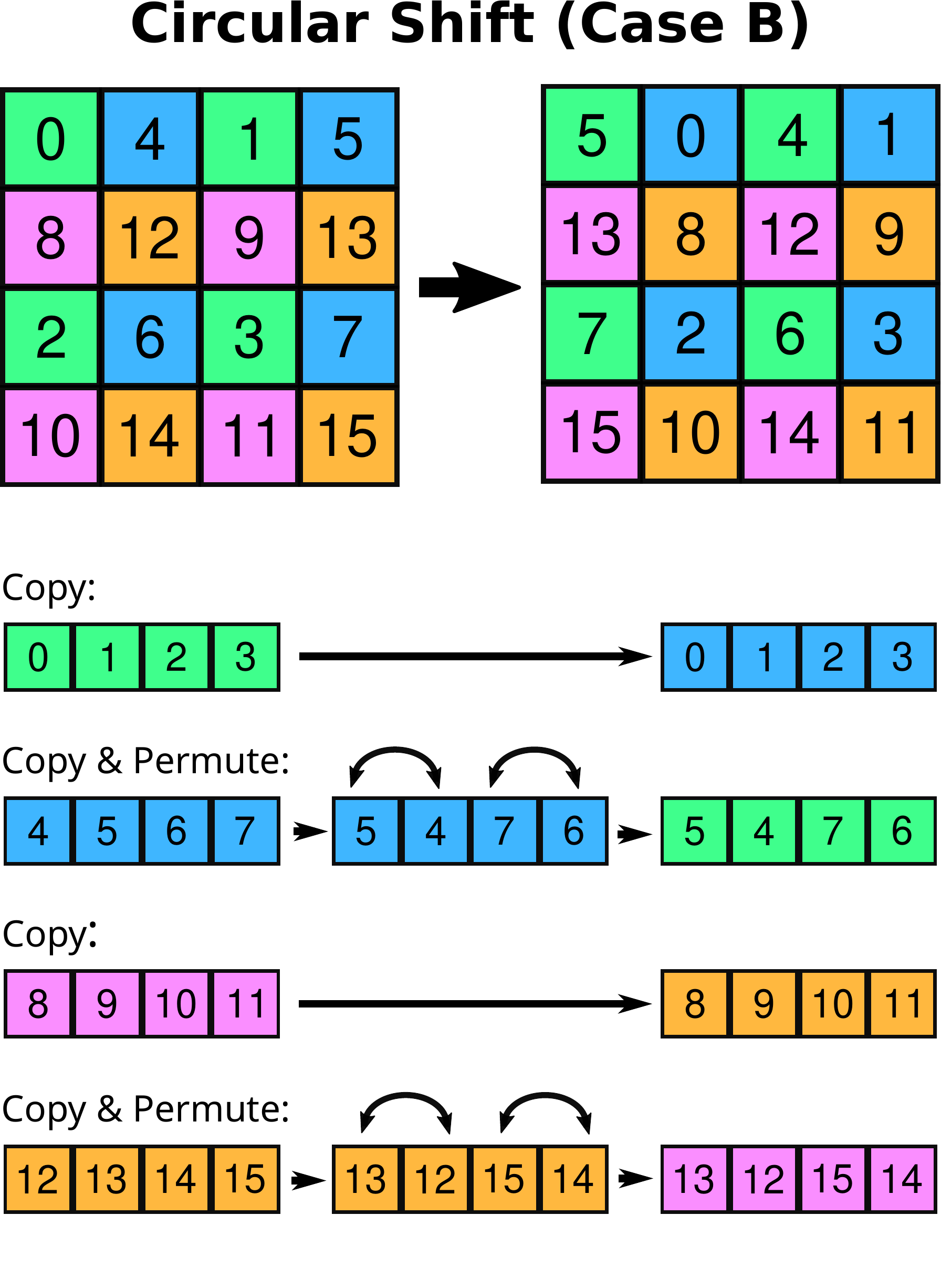}
    \end{minipage}
    \hfill\hfill
    \begin{minipage}[c]{0.46\linewidth}
        \centering
        \includegraphics[width=\linewidth]{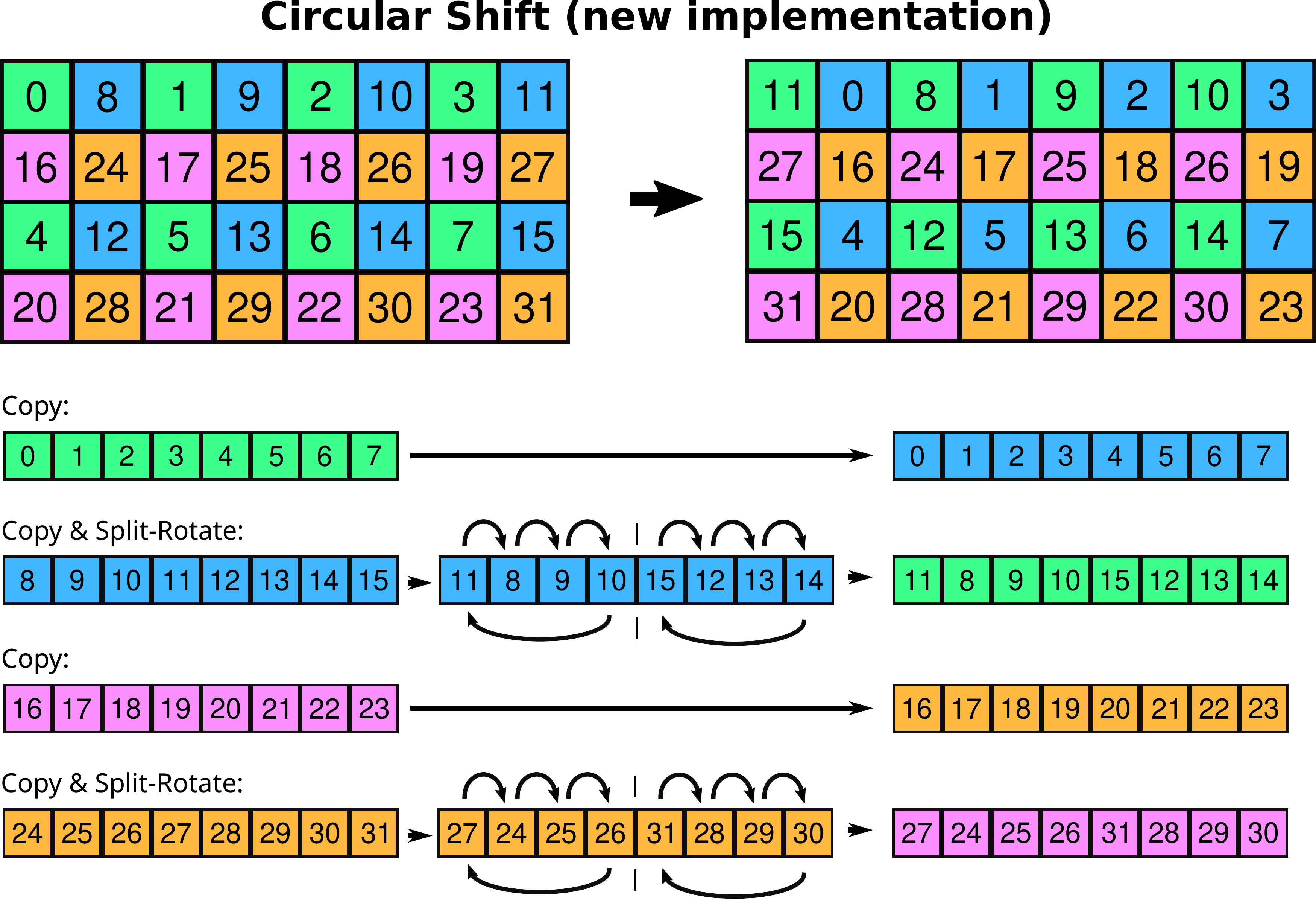}
    \end{minipage}
    \caption{Examples for shifts in a 2d lattice with $m=1$. Case A (\texttt{rotate}) has 
    SIMD layout \{4,1\}, case B (\texttt{permute}) \{2,2\}. In both cases $n=4$. The new implementation \texttt{split\_rotate} (right-most figure, $n=8$) is able to handle a SIMD layout of, e.g., \{4,2\}. The colored boxes denote
    positions in memory, where all boxes with the same color are in contiguous memory.
    At the bottom we display the transformations of the underlying \gridarray{}s. }
    \label{fig:grid_shift}
\end{figure}

On CPUs, mainline Grid is restricted by design to \gridarray{}s of at most 512 bits. This restriction
does not appear directly in the code. Rather, it is an assumption in shift
and stencil operations. 
To explain the origin of this restriction, it is helpful to understand how the lattice sites are mapped to \gridarray{}s. We first introduce some notation.
\begin{itemize}\itemsep-1mm
\item We assume a $d$-dimensional lattice with $V=L_0\cdot\ldots\cdot L_{d-1}$ lattice sites.
\item The degrees of freedom of a single lattice site are given by $m$ numbers. We assume that the data type of these numbers is fixed (e.g., single real, single complex, etc.).
\item A \gridarray contains $n$ numbers of this fixed data type.\footnote{For example, if the \gridarray size is 512 bits, we have $n=4$ for double complex, $n=8$ for double real or single complex, and $n=16$ for single real.}
\item Grid defines an array of $d$ integers $\{n_0,\ldots,n_{d-1}\}$ called ``SIMD layout''.
  These integers are powers of 2 and have to satisfy the condition $n_0\cdot\ldots\cdot n_{d-1}=n$.
\end{itemize}
The mapping then proceeds as follows.
\begin{enumerate}\itemsep-1mm
\item The lattice is decomposed into sublattices containing $n$ lattice sites each.  The number $n_i$ ($i=0,\ldots,d-1$) equals the number of sublattice sites in dimension $i$. Note that Grid performs the decomposition in such a way that adjacent sites of the full lattice are mapped to different sublattices. In fact, the sites of a given sublattice are as far away from one another in the full lattice as possible (see also Fig.~\ref{fig:grid_shift}).
\item A given sublattice is mapped onto $m$ \gridarray{}s. One \gridarray contains one of the $m$ degrees of freedom of all $n$ sublattice sites (see Fig.~\ref{fig:grid_shift} for $m=1$ with real numbers).
\end{enumerate}
We first consider a single Grid process (without MPI communication). Given the decomposition just described, shifts can be done by a few simple methods (see Fig.~\ref{fig:grid_shift}, cases A and B):
\begin{itemize}\itemsep-1mm
\item If no site of a sublattice is part of the lattice boundary in the shift direction, a simple copy of all involved \gridarray{}s is sufficient.
\item Otherwise, in addition to the copy, the elements within a \gridarray must be rearranged.
Mainline Grid can handle two cases:
  \begin{itemize}\itemsep0mm
  \item Case A: A SIMD layout with all entries but one equal to 1.
    Then the Grid function \texttt{rotate} performs the rearrangement.
  \item Case B: Any other SIMD layout than case A, with the restriction that no entry is larger
    than 2. Then the Grid function \texttt{permute} rearranges the sites in the right way.
  \end{itemize}
\end{itemize}
Other SIMD layouts are not supported by mainline Grid, and thus
the maximum SIMD layout for a 4-dimensional lattice is $\{2,2,2,2\}$. For a lattice with SP real numbers this corresponds to a maximum vector register size of 512 bits, which explains the restriction mentioned above.
However, the SX-Aurora with its 16-kbit vector registers requires at least $\{2,4,4,4\}$ (for a DP complex lattice). Therefore, an extension of the shift and stencil algorithms is necessary.

\begin{figure}[t]
        \begin{minipage}{0.45\linewidth}
            \centering
            \includegraphics[width=\linewidth]{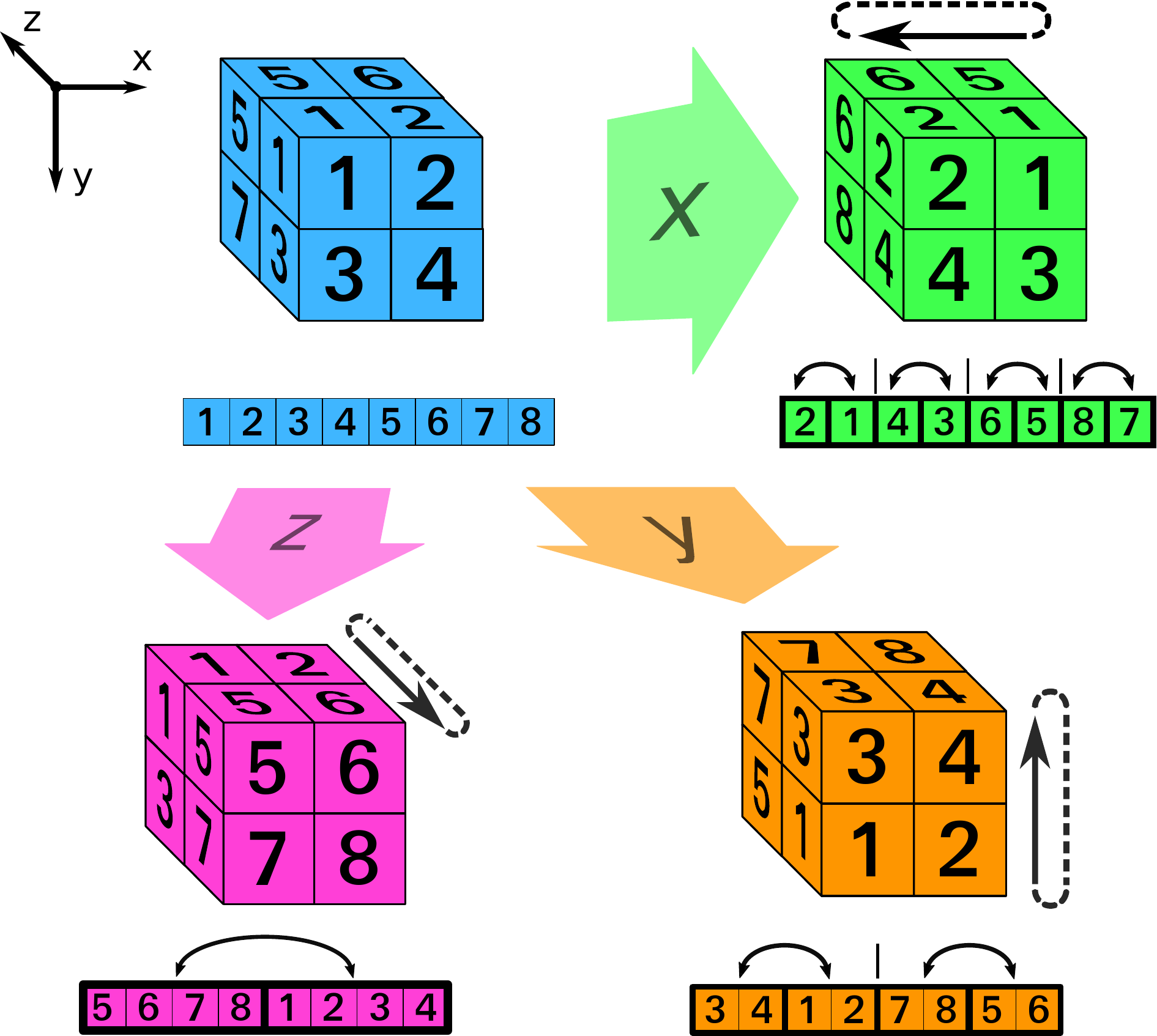}
        \end{minipage}
        \hfill
        \begin{minipage}{0.45\linewidth}
            \centering
            \includegraphics[width=\linewidth]{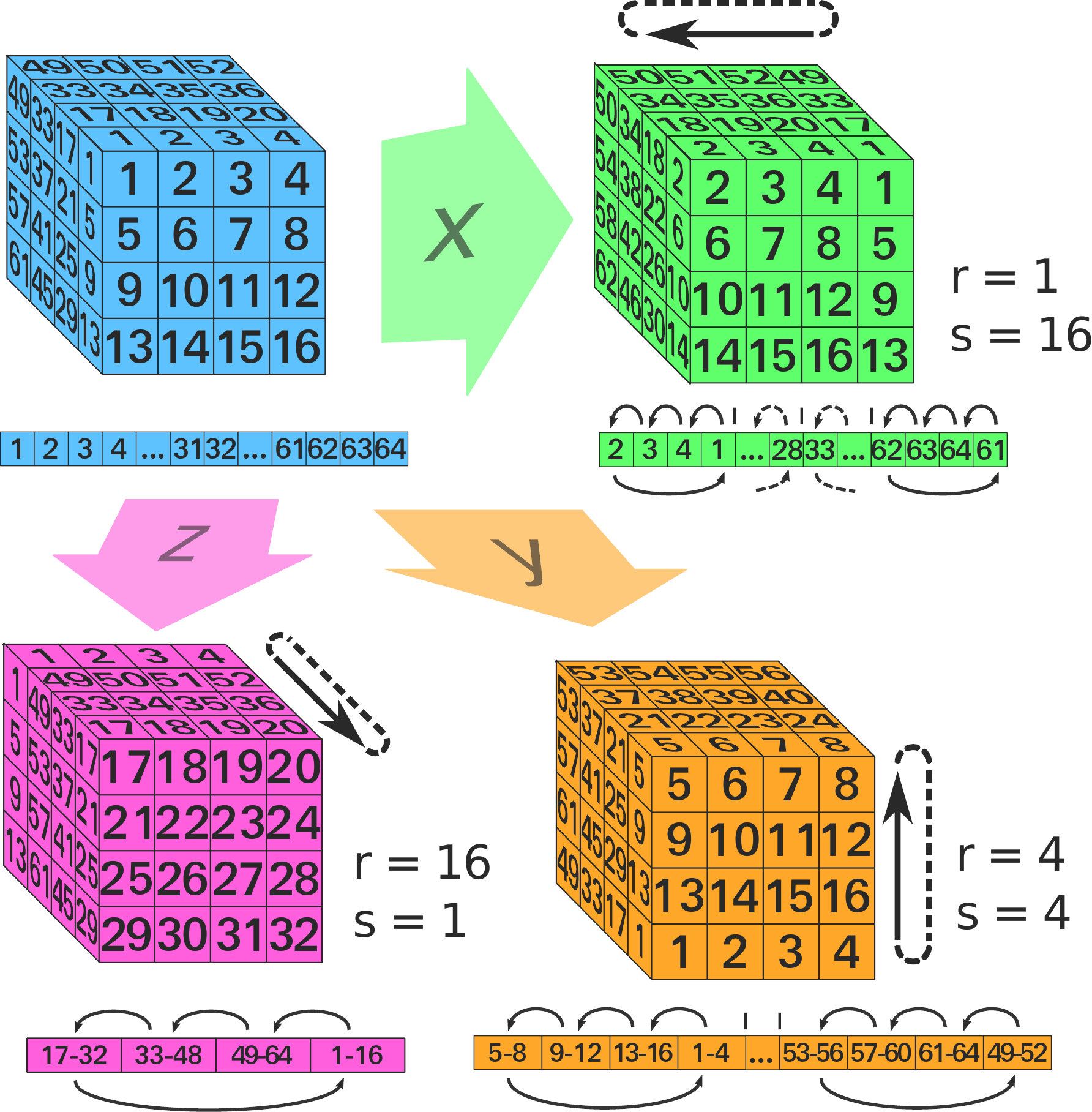}
        \end{minipage}
        \caption{Shifts on a 3d sublattice with $m=1$. Left: A small sublattice with $\text{side length} = 2$ and SIMD layout $\{2,2,2\}$ is shifted by the \texttt{permute} function. Right: A larger sublattice with $\text{side length} = 4$ and SIMD layout $\{4,4,4\}$ is shifted using the \texttt{split\_rotate} function.  The corresponding parameters $ s $ and $ r $ are also shown.}
        \label{fig:3d_shift}
    \end{figure}

The new transformation should have the functionality of \texttt{rotate} and \texttt{permute}, but for an arbitrary SIMD layout. Furthermore, this
operation should be vectorizeable and have similar performance. The new function \texttt{split\_rotate} (shown in the following for the double-precision case) fulfills these aims:
\begin{lstlisting}[texcl]
void split_rotate(double *out, const double *in, int s, int r)
{
    int w = VL/s;         // vector length VL = real DP elements fitting in \gridarray
    for(int i=0; i<VL; ++i)
        out[i] = in[(i+r)%w + (i/w)*w];
}
\end{lstlisting}
The split parameter $ s $ specifies into how many subarrays the \gridarray is split. Then these subarrays are rotated by $ r $. In the case of complex numbers, the input $ r $ must be multiplied by 2. For $ s = 1 $, we obtain mainline Grid's \texttt{rotate} function. The effect of \texttt{split\_rotate} on a 2d lattice is shown in Fig.~\ref{fig:grid_shift}. Examples for shifts on a 3d sublattice by \texttt{split\_rotate} are shown in Fig.~\ref{fig:3d_shift}. \par

We have replaced \texttt{rotate} and \texttt{permute} by \texttt{split\_rotate} in shift and stencil operations and thereby enabled Grid to handle \gridarray sizes of $ 128 \times 2^k $ bits (with $k\in\mathbb N$). Our implementation of \texttt{split\_rotate} is done in generic C/C++ code, which is architecture-independent and thus applicable beyond the SX-Aurora. 

    The algorithm described above works for a single Grid process. When using Grid on multiple nodes (with MPI communication), some \gridarray{}s have to be broken up and partially transferred to the neighboring node in order to perform a lattice shift. In mainline Grid, this is also restricted to the same SIMD layouts as described above. We have enabled the required functionality for larger \gridarray sizes of $ 128 \times 2^k $ bits. However, the implementation still needs to be optimized.

\subsection{Status of porting Grid}
For Grid we have chosen the VE execution model, relying mainly on the ncc compiler and its auto-vectorization capabilities. The progress of porting was slowed down by some compiler and toolchain issues (e.g., Grid compiled with OpenMP only since the summer of 2019). Whereas the issue of enabling larger vector lengths is resolved, the performance of Grid still needs to be tuned. Full MPI support is under ongoing development. We also experimented with Grid and the intrinsics of the clang/LLVM VE compiler, but this option will only become viable once the compiler matures. All sources are available at \cite{Benjamin:2019}, where we forked Grid version 0.8.2.

\subsection{Preliminary performance results}

\begin{figure}[t]
    \begin{subfigure}[t]{.48\linewidth}
      \centerline{\includegraphics[height=42mm]{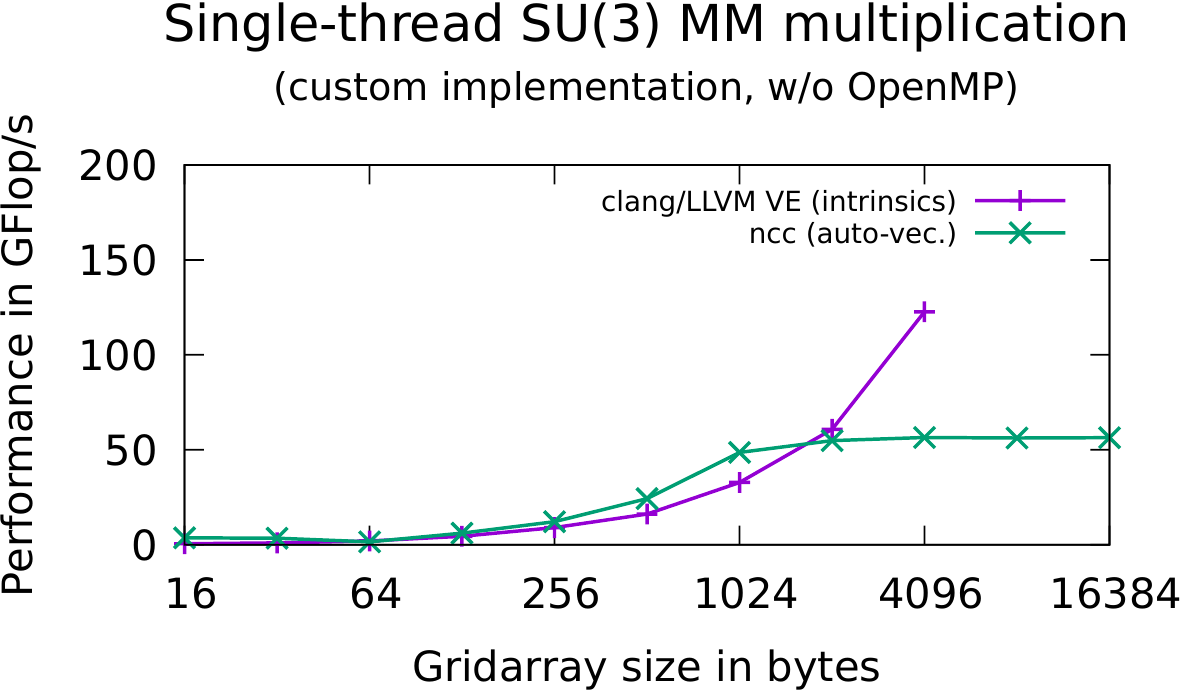}}
        \caption{Single-thread custom SU(3) matrix-matrix multiplication without OpenMP,
        scaling up the size of the \gridarray: intrinsics vs auto-vectorization.}
        \label{fig:register_scale}
    \end{subfigure}
        \hfill
    \begin{subfigure}[t]{.48\linewidth}
        \centerline{\includegraphics[height=42mm]{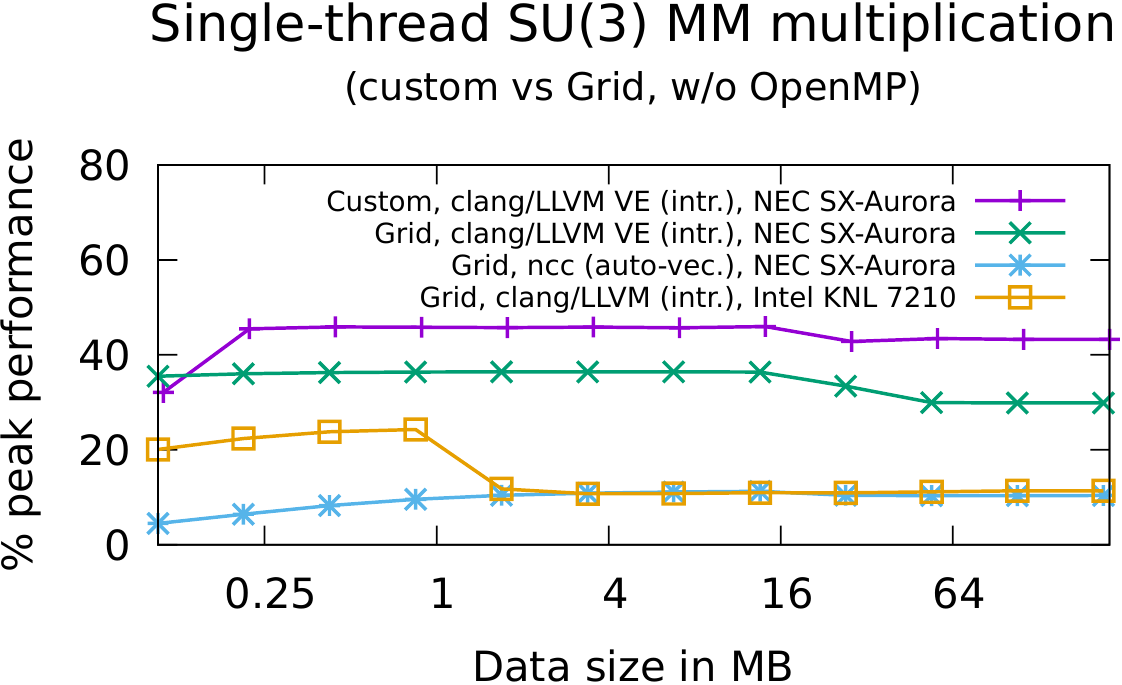}}
        \caption{Single-thread SU(3) matrix-matrix multiplication without OpenMP,
        increasing the lattice size: intrinsics vs auto-vectorization,
        custom vs Grid.}
        \label{fig:su3_lattice_size}
    \end{subfigure}

    \bigskip

    \begin{subfigure}[t]{.48\linewidth}
        \centerline{\includegraphics[height=42mm]{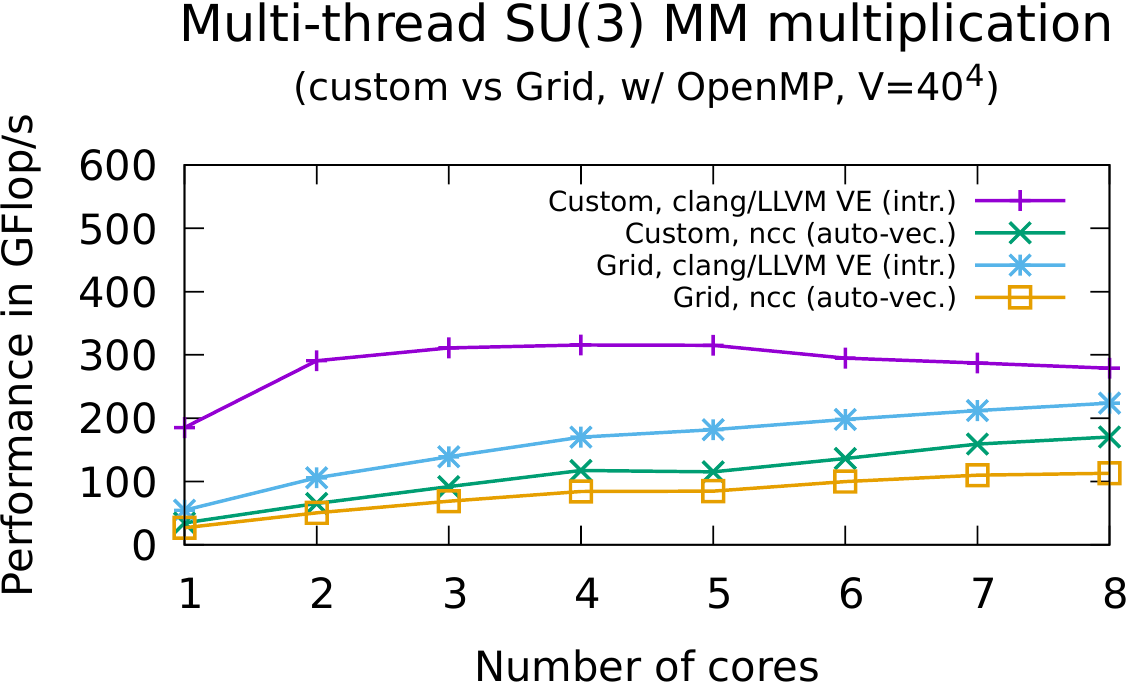}}
        \caption{Multi-thread SU(3) matrix-matrix multiplication using OpenMP:
        intrinsics vs auto-vectorization, custom vs Grid.}
        \label{fig:su3_threads}
    \end{subfigure}
	\hfill
    \begin{subfigure}[t]{.48\linewidth}
        \centerline{\includegraphics[height=42mm]{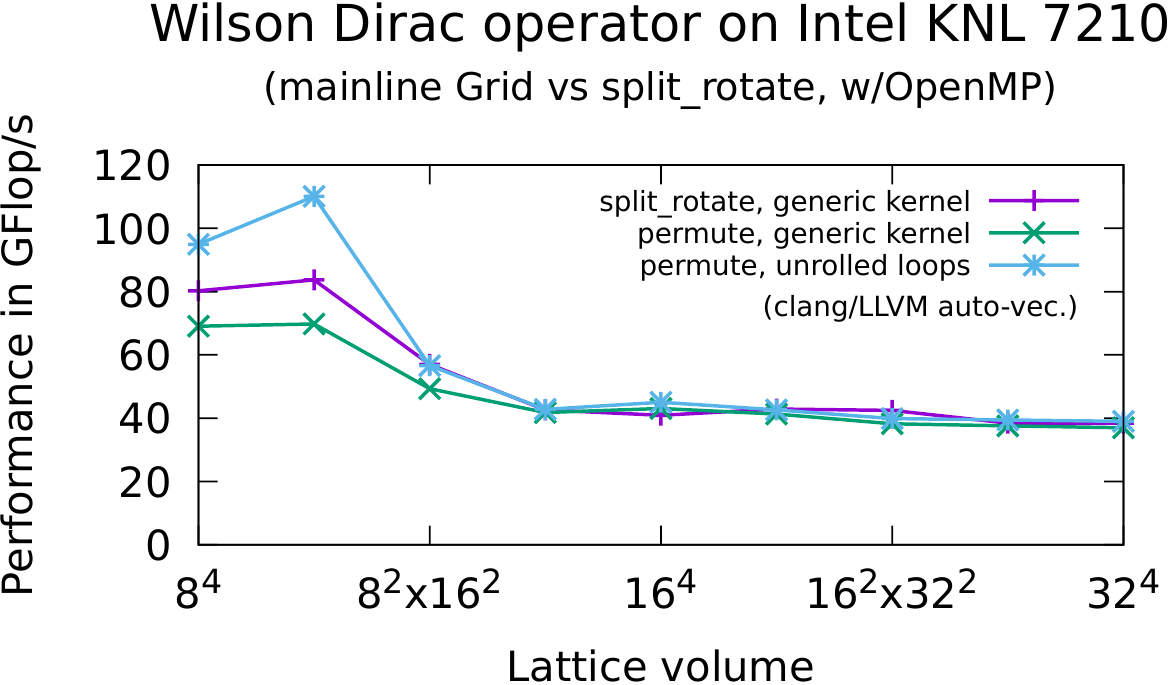}}
        \caption{Application of the Wilson Dirac operator on KNL with OpenMP: mainline Grid implementation (in this case, \texttt{permute} only) vs \texttt{split\_rotate}.}
        \label{fig:KNL}
    \end{subfigure}
    \caption{\label{fig:grid_aurora_bench}Preliminary performance benchmarks. The SX-Aurora benchmarks were performed on a type 10B card using ncc 2.5.1 and clang/LLVM VE 10.0.0. For KNL we used clang/LLVM 5.0.0.}
\end{figure}

In Fig.~\ref{fig:register_scale} we show how the size of the \gridarray{}s affects
the performance of a custom implementation of SU(3) matrix-matrix multiplication.
As in mainline Grid, the data layout in memory is an array of two-element structures $\{\text{re},\text{im}\}$.
Best performance is achieved using clang/LLVM VE intrinsics and when the \gridarray
size is twice the register size ($2\cdot16$ kbit).
The SX-Aurora supports a strided load instruction, which is applied twice to load
the real and imaginary parts into separate registers (strided store is analoguous).
ncc also uses strided load/store, but the performance is significantly worse
due to superfluous copies between register file and LLC.

Figures~\ref{fig:su3_threads} and \ref{fig:su3_lattice_size} show SU(3) matrix-matrix
multiplication (custom implementation vs Grid, both 100\% SIMD efficient) for
two different scaling scenarios: increasing lattice volume using a single thread, and strong thread scaling at constant volume.  Again, the performance of clang/LLVM VE
intrinsics is significantly better than auto-vectorization by ncc due to the 
superfluous copies mentioned above.  For comparison we also show the performance of the Intel
KNL 7210 in Fig.~\ref{fig:su3_lattice_size}.

Figure~\ref{fig:KNL} compares the performance of both shift implementations on a platform where both functions can be called, here Intel KNL 7210 with vector length 512 bits. In the generic version of the Wilson kernel, \texttt{split\_rotate} performs slightly better than \texttt{permute}. Both are surpassed by the hand-unrolled version using \texttt{permute}, which we did not implement for \texttt{split\_rotate}.

\section{Summary and outlook}

We have shown how to modify Grid to deal with larger vector register sizes than the current 512-bit limit and presented performance benchmarks on the SX-Aurora. Work on MPI support is in progress. Once this support is available and further performance optimizations are implemented, the SX-Aurora will be an interesting option for Lattice QCD.

\section*{Acknowledgment}

This work was supported by DFG in the framework of SFB/TRR 55 (project QPACE 4). We thank Erlangen University, Germany, and KEK, Japan, for access to the
SX-Aurora and for support.

\bibliographystyle{JHEP_lat19}
\bibliography{references}

\end{document}